\documentclass[genTeX]{nrc1}   
\usepackage[pctex32]{graphicx}

 \volyear{99}{2005} \received{2005}
\accepted{2005}
\begin{document}

\title{The Uehling correction to the energy levels in a pionic atom}
\author{S. G. Karshenboim}
\address{D. I. Mendeleev Institute for Metrology, St. Petersburg 190005, Russia and
Max-Planck-Institut f\"ur Quantenoptik, Garching 85748, Germany;
e-mail: sek@mpq.mpg.de}
\author{E. Yu. Korzinin}
\address{D. I. Mendeleev Institute for Metrology, St. Petersburg 190005, Russia;
e-mail: korzinin@vniim.ru}
\author{V. G. Ivanov}
\address{Pulkovo Observatory, St. Petersburg 196140,
Russia and D. I. Mendeleev Institute for Metrology, St. Petersburg
190005, Russia; e-mail: vgi@gao.spb.ru}

\shortauthor{S. G. Karshenboim et al.}

\maketitle

\begin{abstract}
We consider a correction to energy levels in a pionic atom induced
by the Uehling potential, i.e., by a free electron
vacuum-polarization loop. The calculation is performed for
circular states ($l=n-1$). The result is obtained in a closed
analytic form as a function of $ Z\alpha$ and the pion-to-electron
mass ratio. Certain asymptotics of the result are also presented.
\end{abstract}
\begin{resume}
Nous ... French version of abstract (supplied by CJP)
   \traduit
\end{resume}

\section{Introduction}

The vacuum polarization effects play an important role in
precision theory of energy levels and wave functions of atoms.
While in conventional atoms such as hydrogen the effects of the
self energy of an electron dominate over the vacuum polarization,
in atoms with a heavier orbiting particle (such as a muon or pion)
the vacuum polarization is responsible for the dominant QED
effects.

In principle, studies of pionic and other exotic atoms involve
effects of the strong interaction. There is a direct interaction
of the pion and the nucleus. Additionally, the finite-nuclear-size
effects are important. However, relations between values of
various contributions depend on values of the nuclear charge $Z$
and of the principal and orbital quantum numbers, $n$ and $l$,
respectively. There are certain situations where the vacuum
polarization dominates. In a number of other cases, it produces a
correction, which is not dominant, but still important.

In this paper we focuss on study of effects of the vacuum
polarization within a relativistic framework. Their leading
contribution to the energy levels is related to so-called Uehling
potential, a correction to the Coulomb potential induced by a free
electronic vacuum loop.

The non-relativistic results in muonic atoms (i.e., the results in
the leading order in $ Z\alpha$) have been known for a while
\cite{pusto} as well as various numerical relativistic
calculations for the conventional (see, e.g., \cite{mohr_pr} and
references therein) and muonic (see, e.g., \cite{borie} and
references therein) atoms.

A relativistic calculation of vacuum polarization effects can be
performed analytically. In particular, various effects induced by
the Uehling potential were studied for the Dirac equation, i.e.,
for an orbiting particle with spin $1/2$, in
\cite{cjp98,cjp98a,cjp01}. The results were obtained in a closed
analytic form as a function of $ Z\alpha$ and a mass ratio of the
orbiting particle and an electron. However, the result for pionic
atoms was not derived because it was unclear how to write
relativistic perturbative series for particle with zero spin.

Here, we extend our approach onto spinless orbiting particles,
such as a pion. Relativistic description of such a particle is
significantly different from that of an electron. A spin-1/2
particle is described by an equation of the form
\[
H\Psi_i=E_i\Psi_i\;,
\]
where the Hamiltonian $H$ is a Hermitian operator and the wave
function $\Psi_i$ is its eigenfunction. The correction to the
energy due to a perturbation by an electrostatic potential $\delta
V$ is of an obvious form
\begin{equation}
\label{shrE} \delta E_i=\langle\Psi^{(0)}_i\vert\delta
V\vert\Psi^{(0)}_i\rangle\;.
\end{equation}
In \cite{cjp98,cjp98a,cjp01} various integrals, which could appear
in calculation of such a martix element over the Uehling potential
in Coulomb bound system were studied.

The equation, that describes a pion, is Klein-Gordon-Fock
equation, which for electrostatic potential is of the form
\[
\left[(E_i-V({\bf r}))^2-{\bf p}^2c^2-m^2c^4\right]\Phi_i({\bf
r})=0\;.
\]
This equation cannot be considered as a problem for eigenfunctions
and eigenvalues of a Hermitian operator. The solutions of this
equation, $\Phi_i$, are not orthogonal and their
normalization is not trivial.

While `exact' solution of the Klein-Gordon-Fock equation with a
non-Coulomb potential has been possible (see, e.g., study of the
problem of falling to the center in \cite{popov} and numerical
calculations in \cite{soff}), the perturbative theory with the
Coulomb field has not been developed.

A simple perturbative series for the Klein-Gordon-Fock equation
has not been known until now and here we apply a recent result
obtained in \cite{kgf_pert}. In particular, the analog of
Eq.~(\ref{shrE}), convenient for analytic calculations, is
 \begin{equation}
  \label{Ebasic}
  \delta E_i=
  \frac{\langle\Phi^{(0)}_i\vert(E^{(0)}_i-V^{(0)})\,\delta V\vert\Phi^{(0)}_i\rangle}
       {\langle\Phi^{(0)}_i\vert(E^{(0)}_i-V^{(0)})\vert\Phi^{(0)}_i\rangle}\;.
  \end{equation}
This equation is valid for any perturbation that can be described
by a certain electrostatic potential and, in particular, for the
Uehling correction.

To calculate the Uehling correction we consider the Coulomb
problem as an unperturbed case. Its eigenvalues and eigenfunctions are
known (see, e.g., \cite{Davidov}):
 \begin{eqnarray}
\Phi^{(0)}_{nl}&=&\frac1r\, R_{nl}(r)\,Y_{lm}(\theta,\varphi)\;,\nonumber\\
R_{nl}(r)&=&N\,\bigl(\beta r\bigr)^{s+1}\exp{\left(-\frac{\beta r}{2}\right)}\,{_1F_1}\bigl(-n+l+1,2s+2,\beta r\bigr)\;,\nonumber\\
E^{(0)}&=&\frac{mc^2}{\sqrt{1+(\frac{ Z\alpha}{\eta})^2}}\;,
 \label{WFKGF}
 \end{eqnarray}
where $Y_{lm}(\theta,\varphi)$ stands for spherical harmonics,
${_1F_1}(a,b;z)$ is the confluent hypergeometric function,
    \begin{eqnarray}
 s&=&-\frac12+\sqrt{\left(l+\frac12\right)^2-( Z\alpha)^2}
  \simeq l-\frac{(Z\alpha)^2}{2l+1}+ \dots\;,\nonumber\\
  \eta&=&n-l+s
  \simeq n-\frac{(Z\alpha)^2}{2l+1}+ \dots\;,\nonumber\\
  \beta&=&\frac{2}{c\,\hbar}\,\frac{( Z\alpha) E^{(0)}}{\eta}
  \simeq \frac{2(Z\alpha)\,mc}{n\,\hbar}\,
  \left(1+\frac{2n-2l-1}{2n^2(2l+1)}\,(Z\alpha)^2 + \dots \right)\;,\nonumber
  \end{eqnarray}
and $m$ is the mass of the orbiting
particle, presumably, a pion. Here, $N$ is a normalization factor,
unspecified in \cite{Davidov}, which is sufficient for our
calculation, since the normalization is unimportant for any
application of Eq.~\ref{Ebasic}.

Our approach can be applied for an arbitrary state. In this paper
we consider only circular states ($l=n-1$), in which the wave
function is of the simplest form
\begin{equation}\label{wfcirc}
  R_{nl}(r)=N\,\bigl(\beta r\bigr)^\eta\exp{\left(-\frac{\beta r}{2}\right)}\;.
\end{equation}
Such states in particular include the $1s$ and $2p$ states. Some
of these states are also of experimental interest: the ground
state ($1s$) is the easiest to create. Meantime, high-$l$ states
are easier for theoretical interpretation because of a substantial
reduction of the strong-interaction effects.

The Uehling potential, which is considered as a perturbation of
the unperturbed problem with the Coulomb potential, is of the form
\cite{Schwinger}
  \begin{eqnarray}\label{dvu}
  \delta V &=&
   -\frac{\alpha( Z\alpha)}{\pi}\int\limits_0^1 dv
  \frac{v^2(1-\frac13v^2)}
       {1-v^2}\,
  \frac{e^{-\lambda\, r}}{r}\;,\nonumber\\
  \lambda&=&\frac{2}{\hbar}\frac{m_ec}{\sqrt{1-v^2}}\;,
  \end{eqnarray}
where $m_e$ is the electron mass.

The integrals needed for matrix element in (\ref{Ebasic}) are very
similar to the ones studied by us previously for a Dirac particle
\cite{cjp98,cjp98a,cjp01}. The result of calculations of the
correction to the energy (\ref{Ebasic}) with potential (\ref{dvu})
over the wave functions from (\ref{WFKGF}) and (\ref{wfcirc}) is
(cf., \cite{cjp98}\footnote{Note, that the integral here,
$K_{abc}(\kappa)$, is slightly different from the one, $
I_{abc}(\kappa,\epsilon)$, used previously in
\cite{cjp98,cjp98a,cjp01}: $K_{a,b,c-2\epsilon}(\kappa)=
  I_{abc}(\kappa,\epsilon)$.})
 \begin{eqnarray}
 \delta E
 &=&
 - \frac{\alpha}{\pi}( Z\alpha)^2
   \frac{mc^2}{\left(\eta^2+( Z\alpha)^2\right)^{3/2}}
   \left[
   \eta\, K_{2,2\eta}(\widetilde{\kappa}_n)+
   \frac{2( Z\alpha)^2}{2\eta-1}\,K_{2,2\eta-1}(\widetilde{\kappa}_n)
   \right] \;,\nonumber\\
 K_{bc}({\kappa})&=&K_{1bc}({\kappa})
 -\frac13K_{2bc}({\kappa})\;,\nonumber\\
  K_{abc}({\kappa})&=&\int_{0}^{1}{dv}\,\frac{v^{2a}}{(1-v^2)^{b/2}}\,
  \left(\frac{{\kappa}\sqrt{1-v^2}}{1+{\kappa}\sqrt{1-v^2}}\right)^{c}
   \label{prom1}\;,\nonumber\\
   \widetilde{\kappa}_n &=&
 \frac{\beta}{2m_e}=
 \frac{( Z\alpha)}{\sqrt{\eta^2+( Z\alpha)^2}}\,\frac{m}{m_e}\;.
 \end{eqnarray}

The results for $K_{abc}({\kappa})$ is known in a closed
analytic form \cite{cjp98}
\begin{eqnarray}
  K_{abc}(\kappa)&=&
  \frac{1}{2}\kappa^c\,
  B\bigl(a+1/2,1-b/2+c/2\bigr)
\\
  &\times &
  {_3F_2}\bigl(c/2,\, c/2+1/2,\, 1-b/2+c/2 ;\;
  1/2,\, a+3/2-b/2+c/2 ;\; \kappa^2\bigr)
 \nonumber\\
  &-&  \frac{c}{2}\,\kappa^{c+1}\,
  B\bigl(a+1/2, 3/2-b/2+c/2\bigr)
  \nonumber\\
  &\times &
  {_3F_2}\bigl(c/2+1,\, c/2+1/2,\, 3/2-b/2+c/2;\;
  3/2,\, a+2-b/2+c/2;\; \kappa^2\bigr)
\;.
  \nonumber
\end{eqnarray}
where ${_3F_2}\bigl(a,b,c;\;d,e;\; z\bigr)$ stands for the
generalized hypergeometric function and $B\bigl(a,b\bigr)$ is the
beta function.

\begin{figure}[phbt]
\begin{center}
{\includegraphics[width=0.45\textwidth]{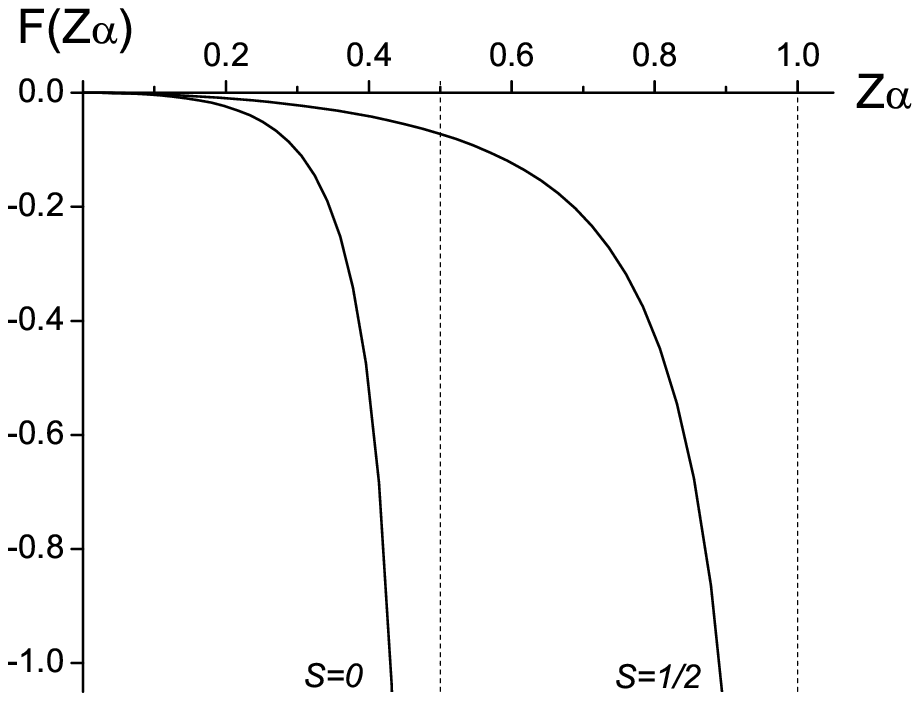}}
{\includegraphics[width=0.442\textwidth]{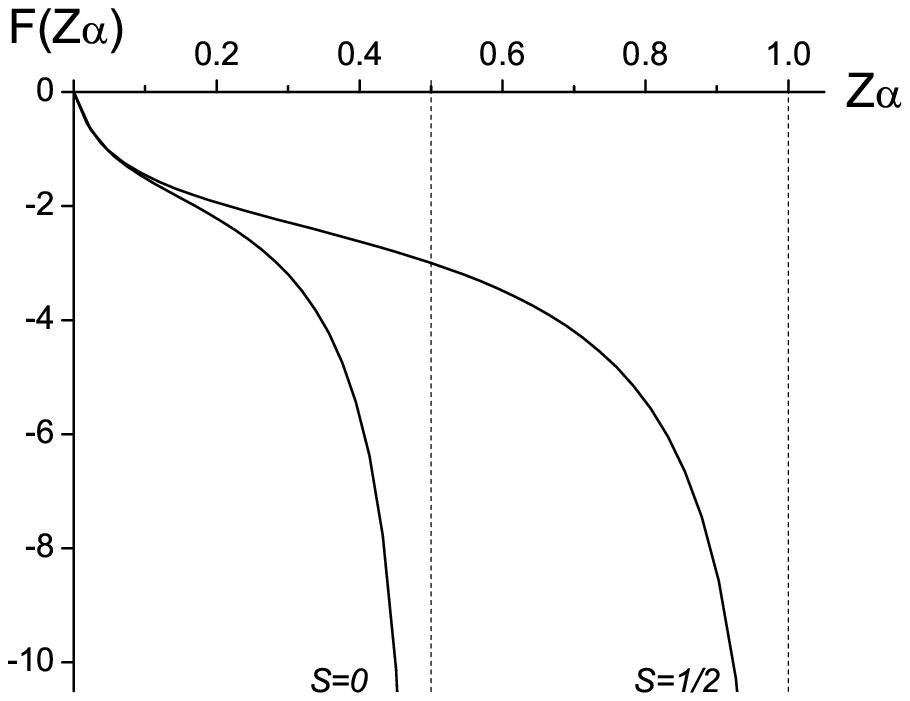}}
\newline
(a)\hspace{6.5cm}(b)\hspace{0.5cm}~
\end{center}
\caption{The Uehling correction to the energy of the $1s$ state
for spin-0 and spin-1/2 particles. The parameters are (a):
$m/m_e=1$ and (b): $m/m_e=250$. \label{pi:fig1}}
\end{figure}

\begin{figure}[phbt]
\begin{center}
{\includegraphics[width=0.45\textwidth]{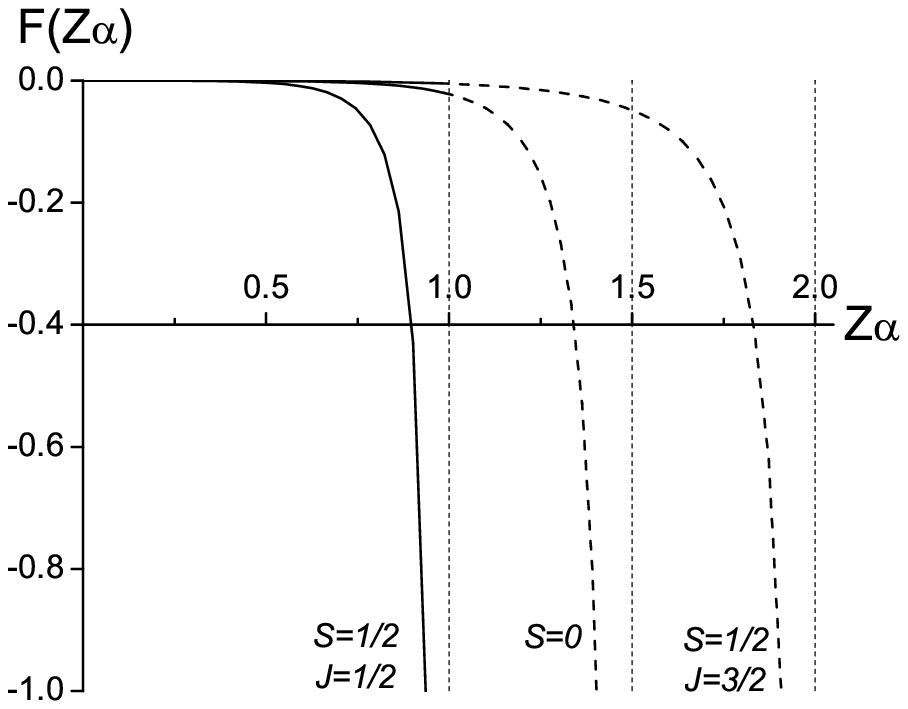}}
{\includegraphics[width=0.445\textwidth]{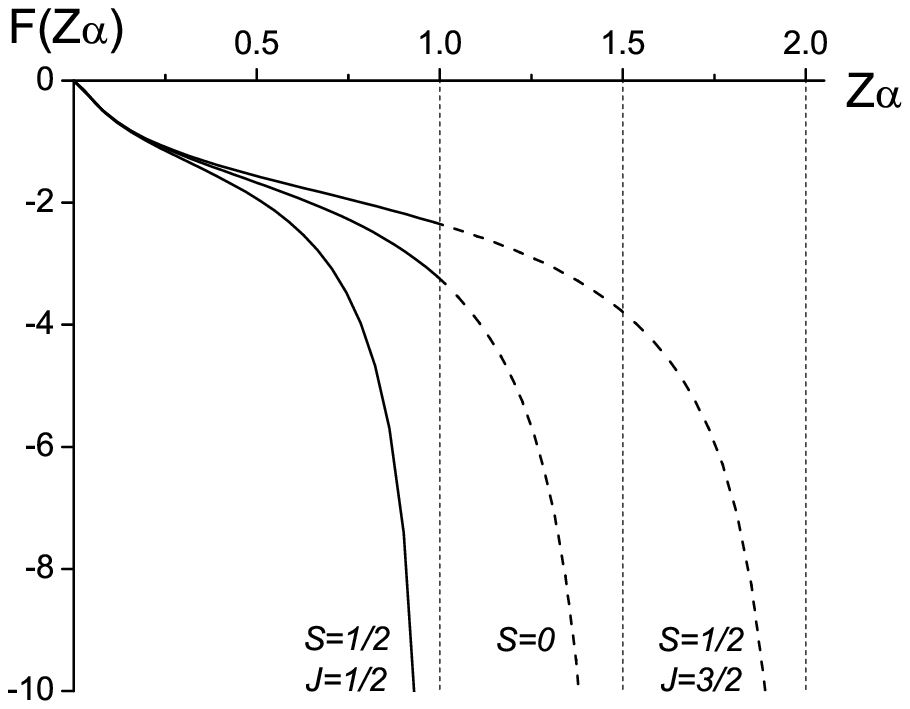}}
\newline
(a)\hspace{6.5cm}(b)\hspace{0.5cm}~
\end{center}
\caption{The Uehling correction to the energy of the $2p$ states
for spin-0 and spin-1/2 particles. The parameters are (a):
$m/m_e=1$ and (b): $m/m_e=250$. \label{pi:fig2}}
\end{figure}

\begin{figure}[phbt]
\begin{center}
{\includegraphics[width=0.45\textwidth]{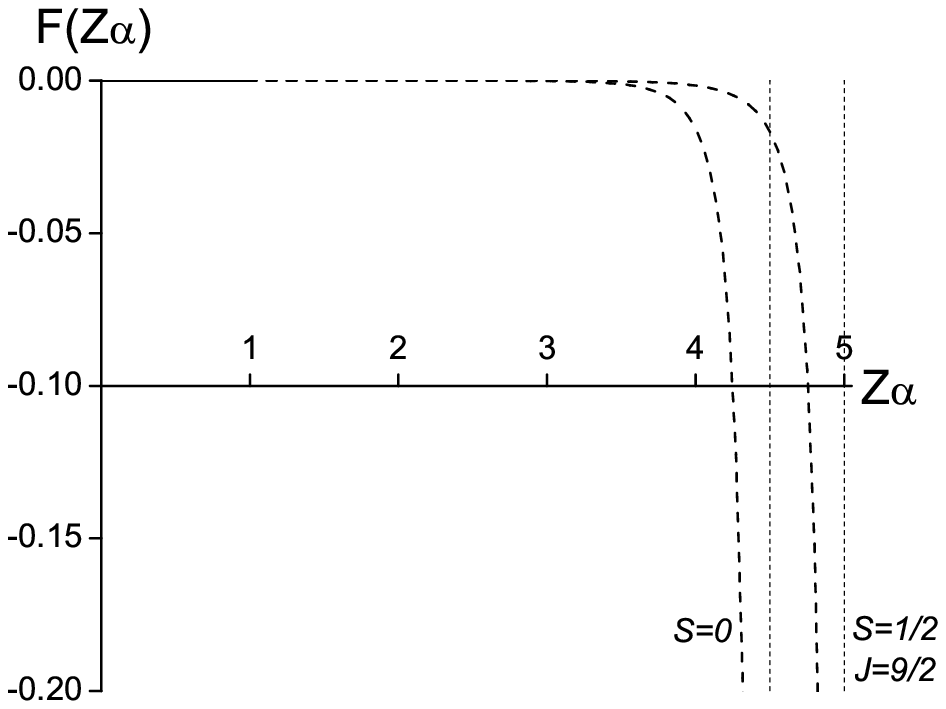}}
{\includegraphics[width=0.44\textwidth]{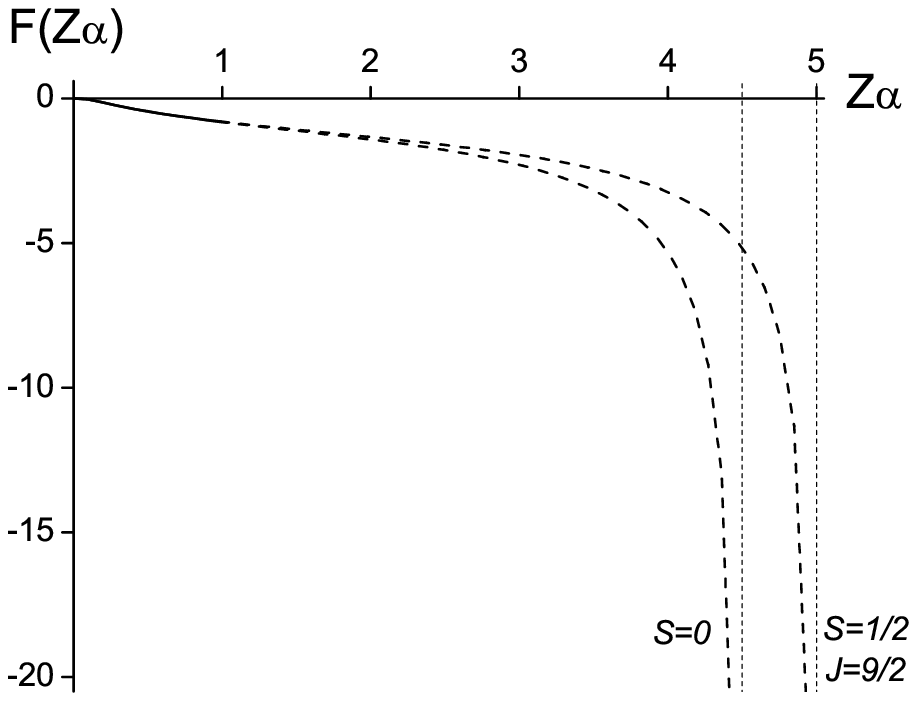}}
\newline
(a)\hspace{6.5cm}(b)\hspace{0.5cm}~
\end{center}
\caption{The Uehling correction to the energy of the $5g$ states
for spin-0 and spin-1/2 particles. The parameters are (a):
$m/m_e=1$ and (b): $m/m_e=250$. \label{pi:fig3}}
\end{figure}

The result for a spin-0 particle is similar that a spin-1/2
particle \cite{cjp98}. We compare the related results for the $1s$
state in Fig.~\ref{pi:fig1}, where we present results for two
values of the mass of the orbiting particle, namely, for $m=m_e$
and $m=250\,m_e$ (for the latter we chose the $m$ value to be
somewhere in between $m_\mu$ and $m_\pi$). The correction is
presented in terms of a dimensionless function $F( Z\alpha)$
defined as
\[
\delta E =\frac{\alpha}{\pi}\,\frac{(Z\alpha)^2mc^2}{n^2}\; F(
Z\alpha)\;.
\]
The similar plots for $2p$ and $5g$ states are presented in
Figs.~\ref{pi:fig2} and \ref{pi:fig3}. We note that the results
have different ranges of convergence, but for physical values
$Z\alpha<1$ the results for excited states are very close
to each other.

The most important difference is related to the range of values
for the nuclear charge $Z$ for which the results, derived here for
a point-like nucleus, can be applied. In principle, it is not
necessary that a correction is well defined for the same range of
parameters as the leading term is. For instance, in a Dirac atom
with a point-like nucleus the hyperfine interval (which can be
considered as a perturbative correction to the electron energy
levels due to an interaction with the nuclear spin) for the $1s$
state is well defined for a narrower range ($Z\alpha<\sqrt{3}/2$)
than the leading term, namely, the Coulomb energy. Nevertheless,
the Uehling correction for both spin-0 and spin-1/2 particles is
well defined in the same $Z$ range as the related Coulomb term. In
particular, for the $1s$ state of a Klein-Gordon particle the
condition is $Z\alpha<1/2$, while for a Dirac particle it is $
Z\alpha<1$. For the hyperfine splitting and the $g$ factor of a
Dirac particle the situation is similar
--- the Uehling correction is valid for the same range of
$Z\alpha$ as the Dirac term (for the reference to the Uehling
correction in the $1s$ state see \cite{cjp98} for the Lamb shift,
\cite{cjp98a} for the hyperfine splitting and \cite{cjp01} for the
$g$ factor of a bound electron).

The result we obtained above (\ref{prom1}) contains two free
parameters: $ Z\alpha$ and
\[
 \kappa_n=\frac{ Z\alpha}{n}\,\frac{m}{m_e}\,,
 \]
 where
 \[
 \widetilde{\kappa}_n =
 \frac{n}{\sqrt{\eta^2+(Z\alpha)^2}}\,\kappa_n
 \simeq \kappa_n \times\left(1+(Z\alpha)^2\,\frac{2n-2l-1}{2n^2(2l+1)}+ \dots\right)\;.\nonumber
 \]
In a pionic atom
\[
\kappa \simeq 2\,\frac{Z}{n}\;,
\]
while in other mesonic atoms this ratio can be even higher.

The general expression is rather complicated and we present here
some useful asymptotics applying an expansion in either of these
two parameters. First we consider a case of $ Z\alpha\ll1$ which
allows a non-relativistic expansion. The result is found to be
\begin{eqnarray}
 \delta E
  =
  &-& \frac{\alpha}{\pi} \frac{(Z\alpha)^2mc^2}{n^2}\,
  \Biggl\{K_{2,2n}(\kappa_n)
  +
 \frac{(Z\alpha)^2}{n(2n-1)}
  \Biggl[
  \frac1{\kappa_n}\,K_{3,2n+1}(\kappa_n)
  \nonumber\\
  &+&
  \left(\frac3{2n}-1\right)\,K_{2,2n}(\kappa_n)
  -2n\,L_{2,2n}(\kappa_n)
  +2K_{2,2n-1}(\kappa_n)
  \Biggr]
  \Biggr\}
 +\dots\;, \label{137}
 \end{eqnarray}
where
 \begin{eqnarray}
 L_{bc}(\kappa)&= &\frac{\partial}{\partial
 c}K_{bc}(\kappa)\nonumber\\
 &=&
 \int_{0}^{1}{dv}\,\frac{v^{2}(1-\frac13\,v^2)}{(1-v^2)^{b/2}}\,
  \left(\frac{\kappa\sqrt{1-v^2}}{1+\kappa\sqrt{1-v^2}}\right)^{c}
  \ln\left(\frac{\kappa\sqrt{1-v^2}}{1+\kappa\sqrt{1-v^2}}\right)
  \;.\nonumber
 \end{eqnarray}
Various asymptotics of the integral above were studied in
\cite{cjp98a,ejp98} (see also below).

An alternative case with $\kappa_n \gg1$ is realized for low-lying
states in medium-Z and heavy pionic atoms. The result is of the
form
\begin{eqnarray}
 \delta E
& =&
 - \frac{\alpha}{\pi}\,
   \frac{( Z\alpha)^2mc^2}{\left(\eta^2+( Z\alpha)^2\right)^{3/2}}
   \Biggl[
   \left(\eta+\frac{2( Z\alpha)^2}{2\eta-1}\right)
   \left(\frac23\ln(2\widetilde{\kappa}_n)+\frac23\psi(1)-\frac59 \right)\nonumber\\
  & -&\frac23\,\eta\,\psi(2\eta)-\frac43\,\frac{( Z\alpha)^2}{(2\eta-1)}\,\psi(2\eta-1)
   +\frac{\pi}2\,\frac{\eta^2+( Z\alpha)^2}{\widetilde{\kappa}_n}
\nonumber\\
   &-&\frac{\eta}{2}\,
   \frac{\eta(2\eta+1)+2( Z\alpha)^2}{\widetilde{\kappa}_n^2}
   +\dots
   \Biggr]\;,
\label{kkkkk}
   \end{eqnarray}
where $\psi(x)=\frac{\partial}{\partial x} \ln(\Gamma(x))$ is the
logarithmic derivative of the Euler gamma function.

To check our calculations we have examined a double asymptotic
behavior for the case $ Z\alpha\ll1$ and $\kappa_n\gg1$. The
results derived from (\ref{137}) and (\ref{kkkkk}) agreed. The
result for the double asymptotics\footnote{Compare the
non-relativistic limit (i.e., the leading term in $(Z\alpha)$) to
the related result in \cite{cjp98} (corrected accordingly to
\cite{cjp01}) and \cite{soto}. We are grateful to J. Soto for
finding a misprint in sign in one of terms in (\ref{misprinted}).}
\begin{eqnarray}\label{misprinted}
 \delta E
  &=&
   -\frac{\alpha}{\pi}\, \frac{( Z\alpha)^2mc^2}{n^2}\Biggl\{\left[
   \frac23\ln(2\kappa_n)+\frac23\bigl(\psi(1)-\psi(2n)\bigr)-\frac59
+\frac{\pi\,n}{2\kappa_n}
  -\frac{n(2n+1)}{2\kappa_n^2}+\dots\right]
  \nonumber\\
  & + &\frac{( Z\alpha)^2}{n(2n-1)}
  \Biggl[
  \left(\frac3{2n}+1\right)
  \left(\frac23\ln(2\kappa_n)+\frac23\,\psi(1)-\frac59\right)
  +\frac1{3n}
\nonumber\\
 &-&\frac23
  \left[\left(\frac3{2n}-1\right)\,\psi(2n)+2\,\psi(2n-1)-2n\,\psi^\prime(2n)\right]
+\frac{4n^2+6n-1}{4\kappa_n^2}
  \Biggr]+\dots
  \Biggr\} + \dots\nonumber
  \end{eqnarray}
is more transparent than the single-parameter expansions
(\ref{137}) and (\ref{kkkkk}).

Other asymptotic cases of interest are related to $\kappa_n\ll1$.
These limit is approached at a pionic atom for high $n$. For
arbitrary $Z\alpha$ the expansion is
\begin{eqnarray}
 \delta E &=&
 -\frac{\alpha}{\pi}\frac{(Z\alpha)^2mc^2}{3(\eta^2+(Z\alpha)^2)^{3/2}}\,
 \biggl[
 \frac{2\eta+1}{2\eta-1}(Z\alpha)^2B\bigl(5/2,\eta-1/2\bigr)
 \,\widetilde{\kappa}_n^{2\eta-1}
 \nonumber\\
 &+&(\eta+1)(\eta-2(Z\alpha)^2)\,
 B\bigl(5/2,\eta\bigr)\,\widetilde{\kappa}_n^{2\eta}\,
 +\ldots\, \biggr]\,
 \end{eqnarray}
while for small $Z\alpha$ we arrive at
 \begin{eqnarray}
 \delta E &=&
 -\frac{\alpha}{\pi}\,\frac{(Z\alpha)^2mc^2}{3n^2}\,
 \biggl\{
 (n+1)B\bigl(5/2,n\bigr)\kappa_n^{2n}
 - n(2n+3)B\bigl(5/2,n+1/2\bigr)\kappa_n^{2n+1}
 +\ldots\,
 \\
 &+&\frac{(Z\alpha)^2}{n(2n-1)}\,
 \biggl[
 (2n+1)B\bigl(5/2,n-1/2\bigr)\kappa_n^{2n-1}
 \nonumber\\
 &-&n(n+1)\biggl(
 2\ln(\kappa_n)+\psi(n)-\psi(n+3/2)+\frac{8n^2-4n-3}{2n^2}
 \nonumber\\
 &+&
 \frac{1}{(n+1)(2n+3)}
 \biggr)B\bigl(5/2,n\bigr)\kappa_n^{2n}
 +\ldots\,
 \biggl]\biggr\}\,.\nonumber
 \end{eqnarray}

Above we considered the Uehling correction to the energy of the
circular states in a pionic atom and found the general expression
and its various asymptotics. The same approach can be applied for
any state. The result for low $Z$, where the non-relativistic
effects dominate, is similar to that for a bound Dirac particle.
At higher $Z$ the correction is different and, in particular, its
has a different range of applicability (see, e.g., Figs. 1--3).
More detail will be presented elsewhere.

\section*{Acknowledgements}

This work was supported in part by the RFBR under grant
03-02-04029 and by DFG under grant GZ 436 RUS 113/769/0-1. Part of
the work was done during visits by EK to Max-Planck-Institut f\"ur
Quantenoptik and we thank MPQ for their hospitality. We are
grateful to Aleksander Milstein for useful discussions.

\end{document}